\documentclass[preprintnumbers,amsmath,amssymbm,prl]{revtex4}
\usepackage{epsfig}
\usepackage{graphicx}
\usepackage{amssymb}

\begin{document}
\title{Asymptotic spectrum of the oblate spin-weighted spheroidal harmonics: a WKB analysis}
\author{Shahar Hod}
\address{The Ruppin Academic Center, Emeq Hefer 40250, Israel}
\address{ }
\address{The Hadassah Institute, Jerusalem 91010, Israel}
\date{\today}

\begin{abstract}
\ \ \ Spin-weighted spheroidal harmonics play a central role in the
mathematical description of diverse physical phenomena, including
black-hole perturbation theory and wave scattering. We present a
novel and compact derivation of the asymptotic eigenvalues of these
important functions. Our analysis is based on a simple trick which
transforms the corresponding spin-weighted spheroidal angular
equation into a Schr\"odinger-like wave equation which is amenable
to a standard WKB analysis.
\end{abstract}
\bigskip
\maketitle


Spin-weighted spheroidal harmonics $S(\theta;c)$ have attracted much
attention over the years from both physicists and mathematicians.
These special functions are solutions of the angular differential
equation \cite{Flam,Teu,Abram}
\begin{equation}\label{Eq1}
{1\over {\sin\theta}}{\partial \over
{\partial\theta}}\Big(\sin\theta {{\partial
S}\over{\partial\theta}}\Big)+\Big[c^2\cos^2\theta-2c
s\cos\theta-{{(m+s\cos\theta)^2}\over{\sin^2\theta}}+s+A\Big]S=0\  ,
\end{equation}
where $\theta\in [0,\pi]$ and $c\in\mathbb{Z}$. When
$c\in\mathbb{R}$, the case we shall study in this Letter, the
eigenfunctions are called oblate. The parameters $m$ and $s$ are the
azimuthal harmonic index and the spin-weight of the wave field,
respectively \cite{Teu}. These quantum numbers can assume integer or
half-integer values \cite{Teu}. When $s=0$ the spin-weighted
spheroidal harmonics reduce to the familiar scalar spheroidal
harmonics \cite{Flam,Teu,Abram} which play a central role in the
mathematical description of diverse physical phenomena such as
electromagnetic wave scattering \cite{Asa}, the quantum-mechanical
description of the hydrogen molecular ion $H_2^+$ \cite{Eyr}, and
modern models of nuclear structure \cite{BDB}.

The general (non-zero) spin case was first studied by Teukolsky
\cite{Teu,Ber1,Hod1} in the context of black-hole perturbation
theory \cite{Notebh}; in this case the parameter $s$ denotes the
type of the perturbation-field: $s=0$ for scalar perturbations,
$s=\pm {1\over 2}$ for massless neutrino perturbations, $s=\pm 1$
for electromagnetic perturbations, and $s=\pm 2$ for gravitational
perturbations.

The angular functions $S(\theta;c)$ are required to be regular at
the poles $\theta=0$ and $\theta=\pi$. These boundary conditions
pick out a {\it discrete} set of eigenvalues $\{_s A_{lm}\}$ labeled
by the discrete parameter $l$. In the $c\to 0$ limit the angular
functions become the familiar spin-weighted spherical harmonics with
the well-known angular eigenvalues $_s A_{lm}=l(l+1)-s(s+1)$ [with
$l\geq \text{max}(|m|,|s|)$]. The opposite limit, $c\to\infty$ (with
fixed $m$), was studied in \cite{Flam,Meix} for the $s=0$ case and
in \cite{Breu,Cas,BCC} for the general spin case. It was found that
the asymptotic eigenvalues are given by \cite{Cas,Notepm}:
\begin{equation}\label{Eq2}
_sA_{lm}=-c^2+\beta |c|+O(1)\  ,
\end{equation}
where the function $\beta$ depends on the spin-parameter $s$, the
azimuthal harmonic index $m$, and the spheroidal harmonic index $l$
[see Eqs. (\ref{Eq16})-(\ref{Eq17}) below].

While correct, the derivation of (\ref{Eq2}) presented in \cite{Cas}
for the general spin case is rather complex and lengthy. The aim of
the present Letter is to present an alternative and (much) shorter
derivation of the formula (\ref{Eq2}) which, we believe, is also
very simple and intuitive. The trick is to transform the angular
equation (\ref{Eq1}) into the form of a Schr\"odinger-like wave
equation and then to perform a standard WKB analysis. It proves
useful to introduce the coordinate $x$ defined by \cite{Yang}
\begin{equation}\label{Eq3}
x\equiv\ln\Big(\tan\Big({{\theta}\over{2}}\Big)\Big)\  ,
\end{equation}
in terms of which the angular equation (\ref{Eq1}) becomes a
Schr\"odinger-like wave equation of the form \cite{NoteSch}
\begin{equation}\label{Eq4}
{{d^2S}\over{dx^2}}-US=0\  ,
\end{equation}
where
\begin{equation}\label{Eq5}
U(\theta)=(m+s\cos\theta)^2-\sin^2\theta(c^2\cos^2\theta-2cs\cos\theta+s+A)\
.
\end{equation}
Note that the interval $\theta\in [0,\pi]$ maps into
$x\in[-\infty,\infty]$. The Schr\"odinger-type angular equation
(\ref{Eq4}) is now in a form that is amenable to a standard WKB
analysis.

The effective potential $U(\theta)$ is in the form of an asymmetric
double-well potential \cite{NoteS}: in the $c\to\infty$ limit it has
a local maximum at
\begin{equation}\label{Eq6}
\theta_{\text{max}}={\pi\over 2}+O(sc^{-1})\ \ \ \text{with} \ \ \
U(\theta_{\text{max}})=c^2-\beta c+O(1)\
\end{equation}
and two local minima at
\begin{equation}\label{Eq7}
\theta^{-}_{\text{min}}=\sqrt{{{\beta-2s}\over{2c}}}+O(c^{-3/2})\ \
\ \text{and} \ \ \
\theta^{+}_{\text{min}}=\pi-\sqrt{{{\beta+2s}\over{2c}}}+O(c^{-3/2})
\end{equation}
with
\begin{equation}\label{Eq8}
U(\theta^{\pm}_{\text{min}})=(m\mp s)^2-{1\over 4}(\beta\pm
2s)^2+O(c^{-1})\ .
\end{equation}
Thus, in the $c\to\infty$ limit the two potential wells are
separated by a large potential-barrier of height
\begin{equation}\label{Eq9}
\Delta U\equiv
U(\theta_{\text{max}})-U(\theta^{\pm}_{\text{min}})=c^2+O(c)\to\infty\
\ \ \text{as}\ \ \ c\to\infty\  .
\end{equation}

Regions where $U(\theta)<0$ are characterized by an oscillatory
behavior of the wave-function $S$ (the `classically allowed
regions'), while regions with $U(\theta)>0$ (the `classically
forbidden regions') are characterized by an exponential behavior
(evanescent waves). The `classical turning points' are characterized
by $U=0$. There are two pairs $\{\theta^{-}_{1,2},
\theta^{+}_{1,2}\}$ of such turning points (with
$\theta^-_1<\theta^{-}_{\text{min}}<\theta^-_2<\theta_{\text{max}}<\theta^+_1<\theta^{+}_{\text{min}}<\theta^+_2$),
which in the $c\to\infty$ limit are located in the immediate
vicinity of the two minima $\{\theta^{-}_{\text{min}},
\theta^{+}_{\text{min}}\}$:
\begin{eqnarray}\label{Eq10}
\theta^{-}_{1,2}=\sqrt{{{\beta-2s\pm\sqrt{(\beta-2s)^2-4(m+s)^2}}\over{2c}}}+O(c^{-3/2})\
\ \ \text{and} \ \ \ \nonumber
\\
\theta^{+}_{1,2}=\pi-\sqrt{{{\beta+2s\pm\sqrt{(\beta+2s)^2-4(m-s)^2}}\over{2c}}}+O(c^{-3/2})\
.
\end{eqnarray}

A standard textbook second-order WKB approximation for the
bound-state `energies' of a Schr\"odinger-like wave equation of the
form (\ref{Eq4}) yields the well-known quantization condition
\cite{WKB1,WKB2,WKB3,Iyer,Notehigh}
\begin{equation}\label{Eq11}
\int_{x^{\pm}_1}^{x^{\pm}_2}dx\sqrt{-U(x)}=(N+{1\over 2})\pi\ \ \ ;
\ \ \ N=\{0,1,2,...\}\  ,
\end{equation}
where $x^{-}_{1,2}$ and $x^{+}_{1,2}$ are the turning points [with
$U(x^{\pm}_{1,2})=0$] of the left and right potential wells,
respectively, and $N$ is a non-negative integer. Here we have used
the fact that in the $c\to\infty$ limit the two potential wells are
separated by an infinite potential-barrier [see Eq. (\ref{Eq9})].
Thus, in the $c\to\infty$ limit the coupling between the wells (the
`quantum tunneling' through the potential barrier) is negligible and
the two potential wells can therefore be treated as independent of
each other \cite{WKB1,Zhou,Notetun}.

Using the relation $dx/d\theta=1/\sin\theta$, one can write the WKB
condition (\ref{Eq11}) in the form
\begin{equation}\label{Eq12}
\int_{\theta^{\pm}_1}^{\theta^{\pm}_2}d\theta{{\sqrt{-U(\theta)}}\over{\sin\theta}}=(N+{1\over
2})\pi\ \ \ ; \ \ \ N=\{0,1,2,...\}\  .
\end{equation}
The WKB quantization conditions (\ref{Eq12}) determine the
eigenvalues $\{A\}$ of the associated spin-weighted spheroidal
harmonics in the large-$c$ limit. The relation so obtained between
the eigenvalues and the parameters $c,m,s$ and $N$ is rather complex
and involves elliptic integrals. However, given the fact that in the
$c\to\infty$ limit the turning points $\theta^{-}_{1,2}$ and
$\theta^{+}_{1,2}$ lie in the immediate vicinity of $\theta=0$ and
$\theta=\pi$ respectively [see Eq. (\ref{Eq10})], one can
approximate the integrals in (\ref{Eq12}) by \cite{Noteapp}
\begin{eqnarray}\label{Eq13}
\int_{\theta^{-}_1}^{\theta^{-}_2}d\theta\sqrt{(\beta-2s)c-c^2\theta^2-{{(m+s)^2}\over{\theta^2}}}=(N+{1\over
2})\pi\ \ \text{and}\
\int_{\theta^{+}_1}^{\theta^{+}_2}d\theta\sqrt{(\beta+2s)c-c^2(\pi-\theta)^2-{{(m-s)^2}\over{(\pi-\theta)^2}}}=(N+{1\over
2})\pi\  .\nonumber \\
\end{eqnarray}
Defining $\phi\equiv \theta\sqrt{{{c}\over{\beta-2s}}}$ for the left
potential well and $\phi\equiv
(\pi-\theta)\sqrt{{{c}\over{\beta+2s}}}$ for the right well, one can
write the two quantization conditions (\ref{Eq13}) in a unified and
compact form
\begin{equation}\label{Eq14}
(\beta\mp
2s)\int_{\phi^{\pm}_1}^{\phi^{\pm}_2}d\phi{\sqrt{1-\phi^2-{{b_{\pm}^2}\over{\phi^2}}}}=(N+{1\over
2})\pi\  ,
\end{equation}
where $b_{\pm}\equiv |m\pm s|/(\beta\mp 2s)$ and
$\{\phi^{-}_{1,2},\phi^{+}_{1,2}\}$ are the rescaled turning points
[where the integrands of (\ref{Eq14}) vanish]. Evaluating the
integrals in (\ref{Eq14}) is straightforward, and one finds
\begin{equation}\label{Eq15}
\beta^{\pm}(N)=2(2N+1\pm s +|m\pm s|)\ \ \ ; \ \ \ N=\{0,1,2,...\}\
\end{equation}
for the two quantized spectra $\{\beta^-(N),\beta^+(N)\}$ (which
correspond to the two potential wells).

Note that $|\beta^+(N)-\beta^-(N)|/4=|s+{1\over 2}(|m+s|-|m-s|)|$ is
always an integer, which implies that the two spectra (\ref{Eq15})
are doubly degenerate above some eigenvalue. Thus, defining
\begin{equation}\label{Eq16}
\beta^{\pm}_0\equiv 2(1\pm s +|m\pm s|)\ ,
\end{equation}
we can unify the two spectra (\ref{Eq15}) and write them in the form
of a single compact formula:
\begin{equation}\label{Eq17}
\beta(N)=4N+\min(\beta^{-}_0,\beta^{+}_0)\ \ \ ; \ \ \
N=\{0,1,2,...\}\
\end{equation}
where the spectrum (\ref{Eq17}) is doubly degenerate for $N\geq
|\beta^+_0-\beta^-_0|/4$ \cite{NoteCas}.

\bigskip
\noindent
{\bf ACKNOWLEDGMENTS}
\bigskip

This research is supported by the Carmel Science Foundation. I thank
Uri Keshet, Oded Hod, Yael Oren, Arbel M. Ongo and Ayelet B. Lata
for helpful discussions.


\begin{thebibliography}{99}
\bibitem{Notebh} In the context of black-hole perturbation theory,
the parameter $c$ stands for $a\omega$, where $a$ is the angular
momentum per unit mass of the black hole and $\omega$ is the
(conserved) frequency of the perturbation mode \cite{Teu}.
\bibitem{Notepm} We shall assume without loss of generality that
$c>0$. For $c<0$ one should simply replace $\{s,m\}\to\{-s,-m\}$ in
the final answer [see Eqs. (\ref{Eq16})-(\ref{Eq17}) below].
\bibitem{NoteSch} Note that in the quantum-mechanical terminology
$-U$ stands for ${{2m}\over{\hbar^2}}(E-V)$, where $E, V$, and $m$
are the total energy, potential energy, and mass of the particle,
respectively.
\bibitem{Notehigh} Higher order corrections to the asymptotic eigenvalues [see formula (\ref{Eq17}) below] can
be obtained by using a higher-order WKB analysis \cite{Iyer}.
\bibitem{Notetun} More precisely, the coupling between the two wells (due to the
weak `quantum tunneling' through the high potential barrier)
introduces a correction term of order
$\exp[{-\int_{\theta^-_2}^{\theta^+_1}d\theta\sqrt{U(\theta)}}]$ to
the r.h.s of the WKB quantization condition (\ref{Eq11})
\cite{Zhou,WKB1}. This term is of the order of $e^{-\sqrt{\Delta
U}}\sim e^{-c}\to 0$ [see Eq. (\ref{Eq9})] and is therefore
negligible in the $c\to\infty$ limit \cite{Zhou,WKB1}.
\bibitem{Noteapp} Here we have used the relations $\cos\theta=1-{1\over
2}\theta^2+O(\theta^4)$ and $\sin\theta=\theta+O(\theta^3)$ for
$\theta\simeq 0$ and $\cos\theta=-1+{1\over
2}(\pi-\theta)^2+O((\pi-\theta)^4)$ and
$\sin\theta=\pi-\theta+O((\pi-\theta)^3)$ for $\theta\simeq \pi$.
\bibitem{NoteCas} Note that our discrete resonance parameter $N$ corresponds
to $l-l_{\text{min}}$ in the non-degenerate regime $0\leq
N<|\beta^+_0-\beta^-_0|/4$  and to the integral part of ${1\over
2}(l-l_{\text{min}})+|\beta^+_0-\beta^-_0|/8$ in the degenerate
regime $N\geq |\beta^+_0-\beta^-_0|/4$. Here $l$ is the spheroidal
harmonic index \cite{Cas} with $l_{\text{min}}\equiv
\text{max}(|m|,|s|)$.
\end{thebibliography}
\end{document}